\documentclass[aps,pra,superscriptaddress,twocolumn,showpacs,preprintnumbers,floatfix]{revtex4}
\usepackage{epsfig,graphicx,times}
\usepackage{amstext}
\usepackage{amsmath}            
\usepackage{amssymb}            
\usepackage{graphicx}           
\usepackage{latexsym}
\usepackage{bm}
\begin{document}
\title{Photon-induced tunneling in optomechanical systems}

\author{Xun-Wei Xu}
\affiliation{Institute of Microelectronics, Tsinghua University,
Beijing 100084, China}

\author{Yuan-Jie Li}
\affiliation{Department of Electronic Engineering, Tsinghua
University, Beijing 100084, China}

\author{Yu-xi Liu}
\email[]{yuxiliu@tsinghua.edu.cn}
\affiliation{Institute of
Microelectronics, Tsinghua University, Beijing 100084, China}
\affiliation{Tsinghua National Laboratory for Information Science
and Technology (TNList), Tsinghua University, Beijing 100084, China}

\date{\today}

\begin{abstract}
In contrast to recent studies  [Rabl, Phys. Rev. Lett. \textbf{107}, 063601
(2011); Nunnenkamp {\it et al.}, Phys. Rev. Lett. \textbf{107}, 063602
(2011)] on photon blockade that prevents subsequent photons from
resonantly entering the cavity in optomechanical systems, we study
the photon-induced tunneling that increases the probability of
admitting subsequent photons in those systems. In particular, we
analytically and numerically show how two- or three-photon
tunneling can occur by avoiding single-photon blockade. Our study
provides another way on photon control using a single mechanical
resonator in optomechanical systems.

\end{abstract}

 \maketitle

\section{Introduction}

Optomechanical systems~\cite{Kippenberg, Marquardt} have attracted
extensive attention in the past years because of their potential
application in high-precision measurements and quantum information
processing. To realize these benefits, the mechanical resonator has
to be in its ground state; and the optomechanical radiation-pressure
interaction strength should be bigger than the decay rates of the
cavity field and the mechanical oscillator. The ground state cooling
of the mechanical resonator has been experimentally studied in these
systems (e.g., in Refs.~\cite{Arcizet,Kleckner, Gigan, Schliesser,
TeufelPRL, Thompson, Rocheleau, Teufel475}). Although the strong
coupling is not easily achieved in standard optomechanical
systems, the experimentalists have obtained an effectively
strong coupling, by applying a classical driving field to the cavity
mode, which has led to observations of normal-mode splitting (e.g.,
in Refs.~\cite{Groblacher, Teufel471}) and optomechanically induced
transparency (e.g., in Refs.~\cite{Agarwal, Weis, SafaviNaeini}).
However, the resulting effective coupling resembles two linearly
coupled harmonic oscillators~\cite{Groblacher}, and the coupling
strength is proportional to the square root of the mean cavity
photon number; thus it does not really describe the nonlinear effect
at the single-photon level.

Two recent proposals~\cite{Rabl, Nunnenkamp} showed that
the single-photon effect or photon blockade~\cite{Imamoglu} can
occur when the optomechanical systems are approaching the single-photon strong
coupling~\cite{Gupta,Brennecke,Eichenfield}. This is because the mechanical
resonator parametrically modulates the frequency of the cavity field
and results in the photon-photon
interaction~\cite{Meystre,Mancini,Gong,liao1,liao2,binghe}. If the
strong optomechanical interaction makes the photon-photon
coupling strength bigger than the decay rate of the cavity field,
then the photons can prevent the subsequent photons from resonantly
entering the cavity. The photon blockade has been demonstrated
experimentally in cavity QED systems for
microwave~\cite{Wallraff,Hoffman} and optical~\cite{Kimble,Vuckovic}
photons. Meanwhile, the experimentalists~\cite{Kimble,Vuckovic} also
observed the photon induced tunneling, that is, the probability of
admitting subsequent photons is increased when there is one photon
inside the cavity. Moreover, absorption and emission of resonant
photons in pairs have been observed~\cite{Kubanek, Koch}.

Motivated by recent
works~\cite{Rabl,Nunnenkamp,Imamoglu,Gupta,Brennecke,
Eichenfield,Meystre,Mancini,Gong,liao1,liao2,binghe,Wallraff,Hoffman,Kimble,Vuckovic,Kubanek,
Koch} and also in contrast to the photon
blockade~\cite{Rabl,Nunnenkamp}, here we study photon-induced
tunneling in optomechanical systems. In Sec.~II,  the
theoretical model and the master equation are
introduced. The effect of the mechanical resonator on the mean
photon numbers is discussed. In Sec.~III, the two-photon tunneling is discussed
via the normalized second-order correlation function of the cavity
field. In Sec.~IV, we use the three-photon tunneling as an
example to show the multiphoton tunneling phenomena.
A summary is finally given in Sec.~V.

\section{Theoretical model}

We study an optomechanical system (e.g., review in
Ref.~\cite{Kippenberg}) in which a cavity field is coupled to a
mechanical resonator through the radiation pressure with the
Hamiltonian
\begin{equation}\label{eq:1}
H =\hbar \omega _{0}a^{\dag }a+\hbar \omega _{m}b^{\dag }b +\hbar
Ga^{\dag }a\left( b^{\dag }+b\right),
\end{equation}%
where $a$ ($b$) and $a^{\dagger}$ ($b^{\dagger }$) are the
annihilation and creation operators of the cavity field (mechanical
resonator) with the frequency $\omega_{0}$ ($\omega_{m}$). The
coupling strength between the cavity field and
the mechanical resonator is $G$. By applying a unitary transform
$U=\exp{[-G a^{\dag }a\left( b^{\dag }-b\right)/\omega_{m}]}$ to
Eq.~(\ref{eq:1}), we obtain an effective Hamiltonian $H^{\prime}=
UHU^{\dagger}$ with
\begin{equation}\label{eq:2}
H^{\prime} =\hbar \omega _{m}b^{\dag }b+\hbar \left(\omega
_{0}-\frac{G^{2}}{\omega _{m}}\right)a^{\dag }a-\hbar
\frac{G^{2}}{\omega _{m}} a^{\dag }a^{\dag }aa,
\end{equation}%
which has eigenstates $|n,\widetilde{m}\rangle$ and corresponding eigenvalues
\begin{equation}\label{eq:3}
E_{n,m}=\hbar \left(n\,\omega _{0}-n^{2}\frac{G^{2}}{\omega
_{m}}\right)+\hbar m\, \omega _{m},
\end{equation}%
where $|n,\widetilde{m}\rangle\equiv U|n,m\rangle$. $|n,m\rangle$
represents a state of $n$ photons and $m$ phonons. A unitary transform
does not change the eigenvalues of the Hamiltonian; thus the
eigenvalues of the Hamiltonian in Eq.~(\ref{eq:1}) are the same as
in Eq.~(\ref{eq:3}).

We now assume that the cavity field is driven by a weak probe field
with the frequency $\omega_{c}$ and the coupling strength
$\varepsilon _{c}$. In the rotating reference frame with the unitary
operator $R(t)=\exp[i \omega_{c} a^{\dag }at ]$, the Hamiltonian in
Eq.~(\ref{eq:1}) becomes
\begin{eqnarray}
\widetilde{H} =\hbar \Delta a^{\dag }a+\hbar \omega_{m}b^{\dag }b
 +\hbar G a^{\dag }a\left(b^{\dag }+b\right)
 +i\hbar \varepsilon_{c}\left(a^{\dag}-a\right),
\end{eqnarray}%
where $\Delta=\omega _{0}-\omega _{c}$ is the detuning between the
cavity field and the probe field. When the environmental effect is
taken into account, the dynamical evolution of the reduced density
operator $\rho(t)$ for the cavity field and the mechanical resonator
can be described via the master equation~\cite{Carmichael}
\begin{eqnarray}\label{eq:6}
\frac{d \rho }{d t} &=&\frac{1}{i\hbar }\left[ \widetilde{H},\rho \right] +%
\frac{\gamma}{2}\left( 2a\rho a^{\dag }-a^{\dag }a\rho -\rho
a^{\dag }a\right)  \nonumber\\
&&+\frac{\gamma _{m}}{2}\left( 2b\rho b^{\dag }-b^{\dag }b\rho -\rho
b^{\dag }b\right)  \nonumber\\
&&+\gamma _{m} \bar{n}_{m}\left( b\rho b^{\dag }+b^{\dag }\rho
b-b^{\dag }b\rho -\rho b b^{\dag }\right),
\end{eqnarray}%
with the decay rates $\gamma$ and $\gamma _{m}$ of the cavity field
and mechanical mode.  $\bar{n}_{m}=[\exp(\hbar
\omega_{m}/k_{B}T)-1]^{-1}$ is the thermal phonon number of
the mechanical resonator with the Boltzmann constant $k_{B}$ and
the environmental temperature $T$ of the mechanical resonator. Here,
the frequency of the cavity field is assumed to be high enough; thus
the thermal photon effect can be neglected. In the basis of the states $|n,m\rangle$,
the formal solution of $\rho(t)$ in Eq.~(\ref{eq:6}) can be given by
\begin{equation}\label{eq:7}
\rho (t)=\underset{n,m,n^{\prime},m^{\prime}}{\sum
}\rho_{n,m;n^{\prime},m^{\prime}}(t)| n,m\rangle \langle
n^{\prime}, m^{\prime}|.
\end{equation}

If all elements $\rho _{n,m;n^{\prime },m^{\prime }}(t)$ in
Eq.~(\ref{eq:7}) are given, then any physical quantity of the system
can be obtained. For example, the mean photon number inside the
cavity can be obtained as
\begin{equation}\label{eq:8}
\left\langle n\right\rangle = \left\langle a^{\dag }a\right\rangle ={\rm Tr} [\rho(t) a^{\dag }a]=%
\underset{n,m}{\sum }n\rho _{n,m;n,m}(t)\,.
\end{equation}%
\begin{figure}
\includegraphics[bb=10 5 366 276, width=4 cm, clip]{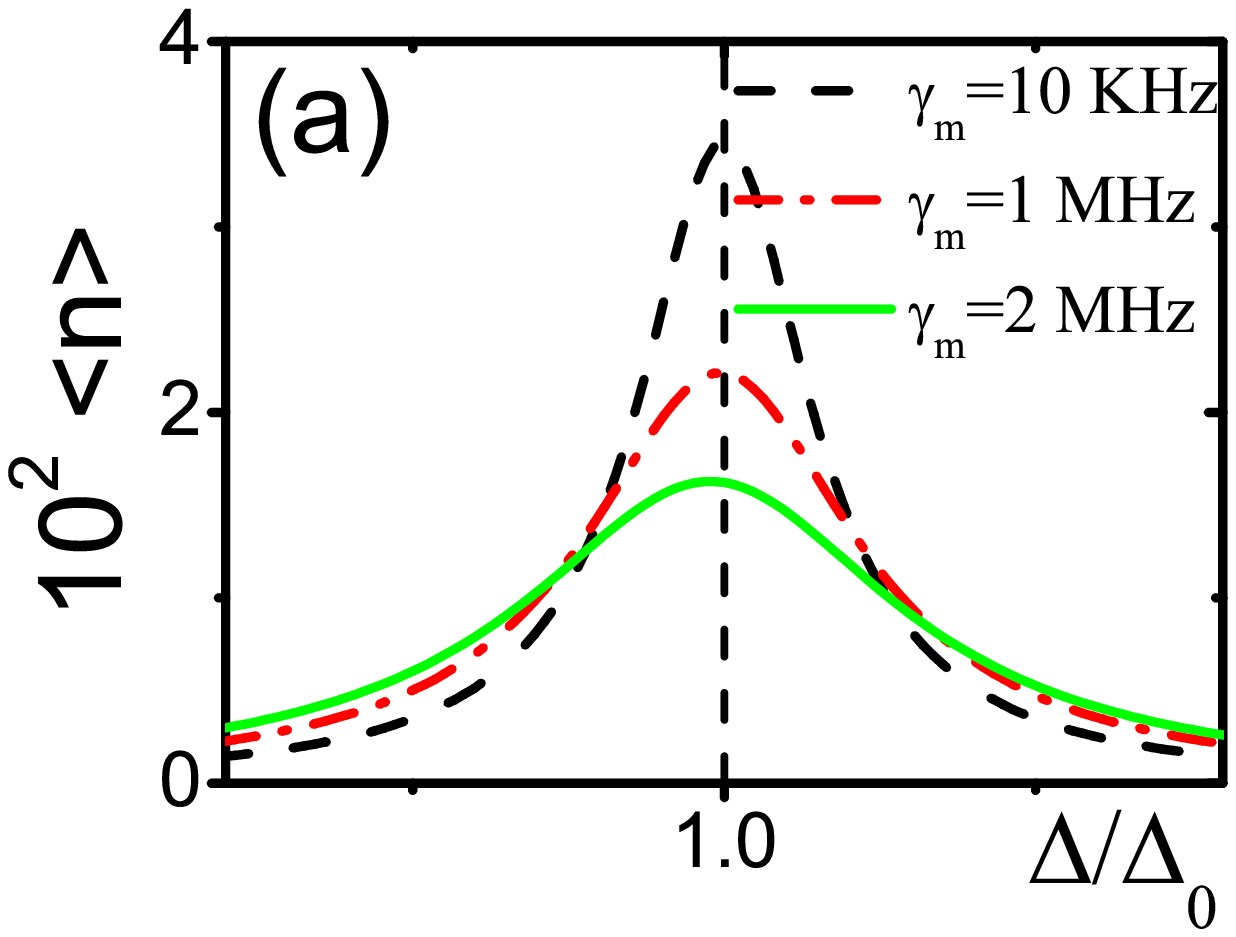}
\includegraphics[bb=10 5 366 276, width=4 cm, clip]{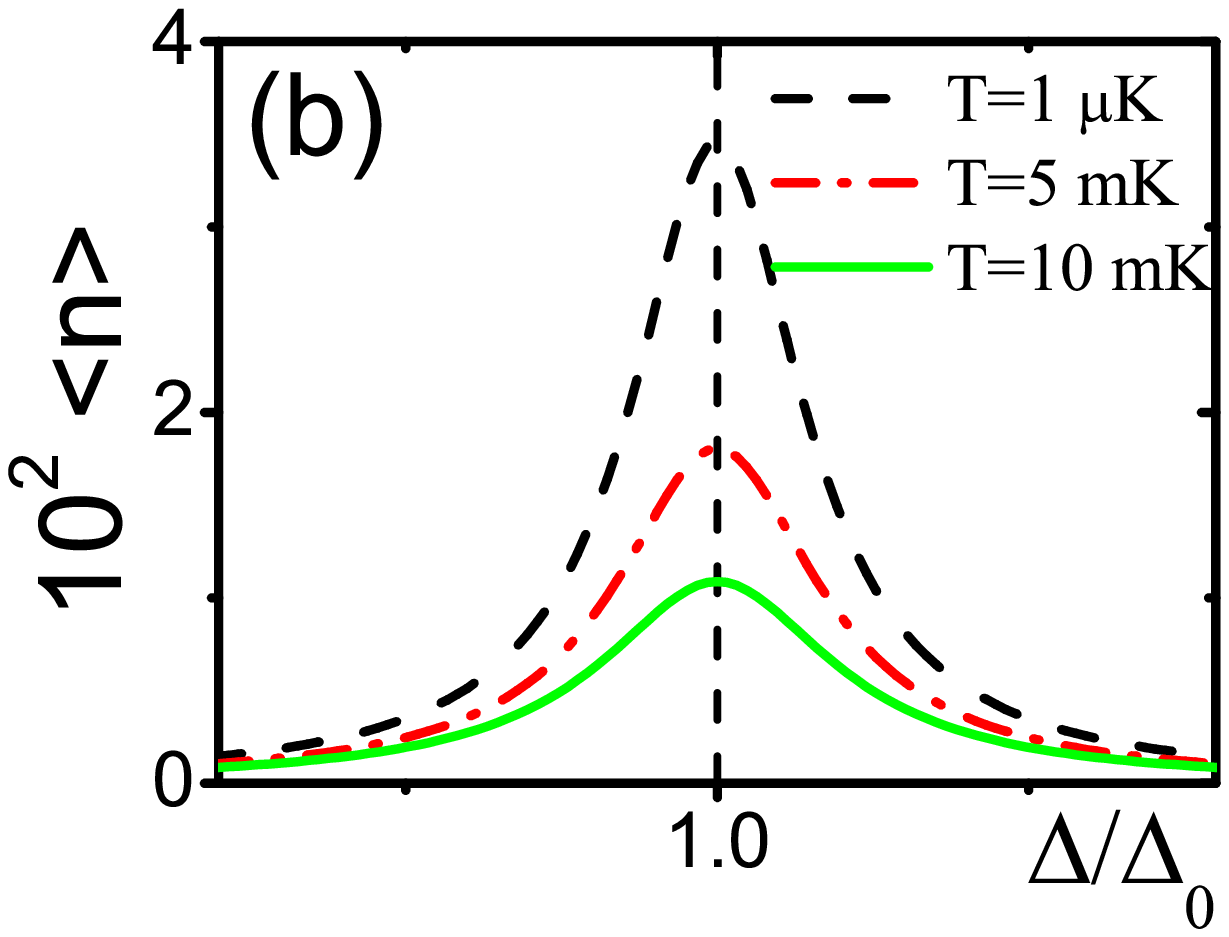}
\caption{(Color online) $\langle n\rangle$ versus $\Delta/\Delta_{0}$ in (a) for different $\gamma_{m}$
at $T=1$ $\mu$K, and (b) for different $T$ at $\gamma_{m}/2\pi=0.01$ MHz. Other parameters are
$\varepsilon_{c}/2\pi=0.01$ MHz, $G/2\pi=2.5$ MHz, $\gamma/2\pi=0.1$
MHz, and $\omega_{m}/2\pi=10$ MHz.} \label{fig1}
\end{figure}
Using Eqs.~(\ref{eq:6})-(\ref{eq:8}), $\langle n\rangle$ is
plotted in the steady state as a function of the normalized detuning
$\Delta/\Delta_{0}$ with $\Delta_{0}=G^{2}/\omega_{m}$ for (i)
different $\gamma_{m}$ in Fig.~\ref{fig1}(a) and (ii) different $T$
in Fig.~\ref{fig1}(b).  We find that $\langle n\rangle$ is maximum
at $(\Delta/\Delta_{0})=1$. As shown in Eq.~(\ref{eq:2}), this is
because the transition frequency of the cavity field from the
one-photon state $|1\rangle$ to the ground state
$|0\rangle$ is shifted to $\omega _{0}-(G^{2}/\omega _{m})$ when the
mechanical resonator is coupled to the cavity field. Moreover,
Fig.~\ref{fig1} shows that when the decay rate $\gamma_{m}$ or the
environmental temperature $T$ of the mechanical resonator increases,
the full width at half maximum in the curve of $\langle n\rangle$
becomes broad for other given parameters. This means that
$\gamma_{m}$ and $T$ significantly affect the lift time of the
cavity photons.

\section{Two-photon tunneling}

The statistical properties of the photons can be characterized by
the normalized $n$th-order correlation function
\begin{equation}\label{eq:5}
g^{(n) }(0) = \frac{\left\langle a^{\dagger n} a^{n}\right\rangle
}{\left\langle a^{\dagger}a\right\rangle^{n}}\equiv
\frac{\text{Tr}(\rho a^{\dagger n} a^{n}) }{\text{Tr}(\rho
a^{\dagger}a)^{n}},
\end{equation}
at the zero time delay. Here,  the $\rho$ of the optomechanical
system can be obtained by solving the master equation in
Eq.~(\ref{eq:6}).

We now study the simplest photon tunneling in pairs. Similar to the
photon blockade, this phenomenon can be characterized by
$g^{\left(2\right) }\left( 0\right)$. When $g^{\left(2\right)
}\left( 0\right)$ is less than one, the photon blockade happens and
single photons come out of the cavity. However, when
$g^{\left(2\right) }\left( 0\right)$ is bigger than one, photons
inside the cavity enhance the resonantly entering probability of
subsequent photons; this can make photons come out of the cavity in
pairs under certain condition. In Refs.~\cite{Kubanek,Majumdar},
the second-order differential correlation function
\begin{eqnarray}
 C^{\left( 2\right)
}\left( 0\right) = \left\langle a^{\dag 2}a^{2}\right\rangle -
\left\langle a^{\dag }a\right\rangle ^{2} \equiv [g^{\left( 2\right)
}\left( 0\right)-1]\left\langle n \right\rangle ^{2},
\end{eqnarray}
at the zero time delay is introduced to characterize the probability
of creating photon pairs simultaneously in the cavity.

\begin{figure}
\includegraphics[bb=10 0 405 300, width=4.1 cm, clip]{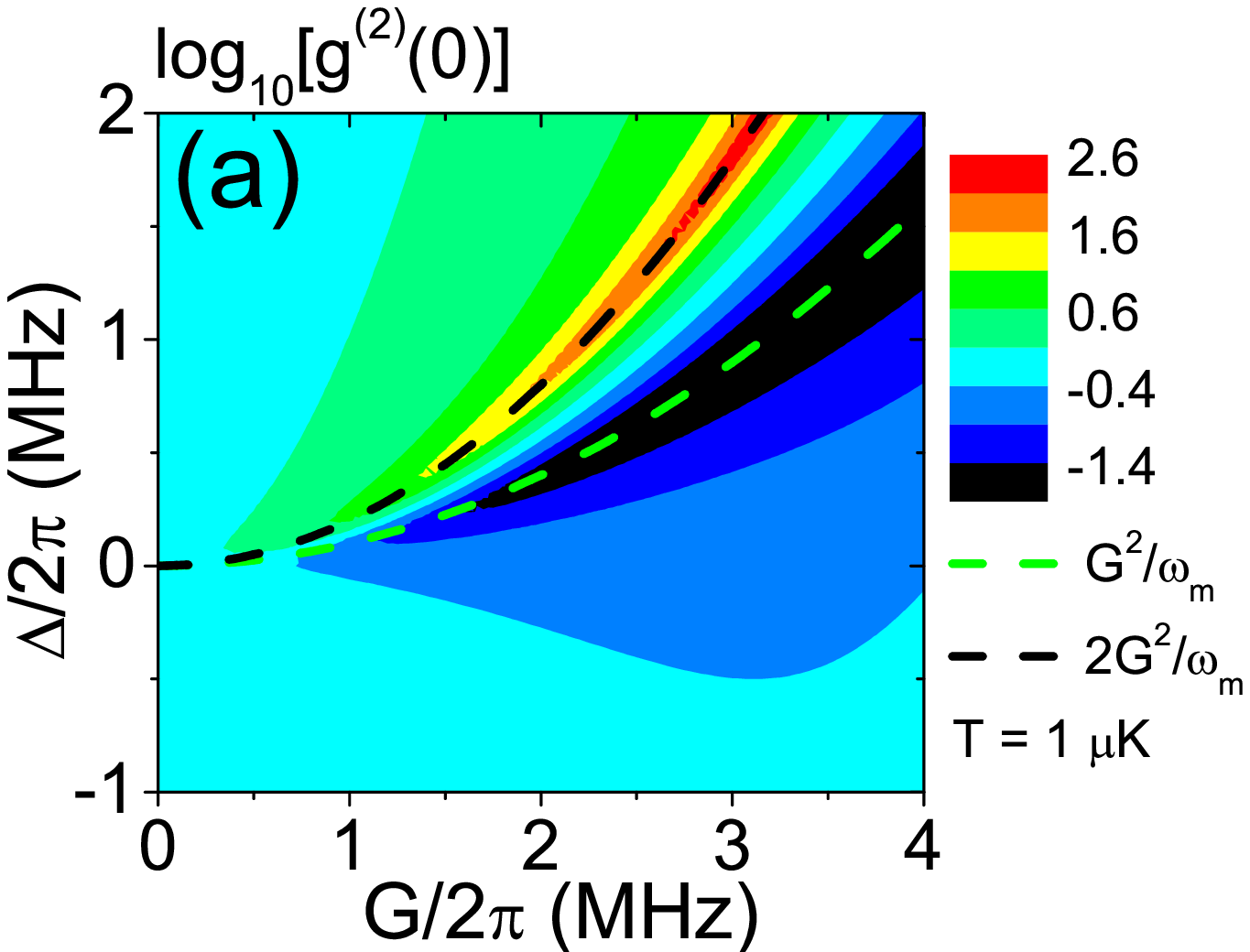}
\includegraphics[bb=0 0 402 297, width=4.1 cm, clip]{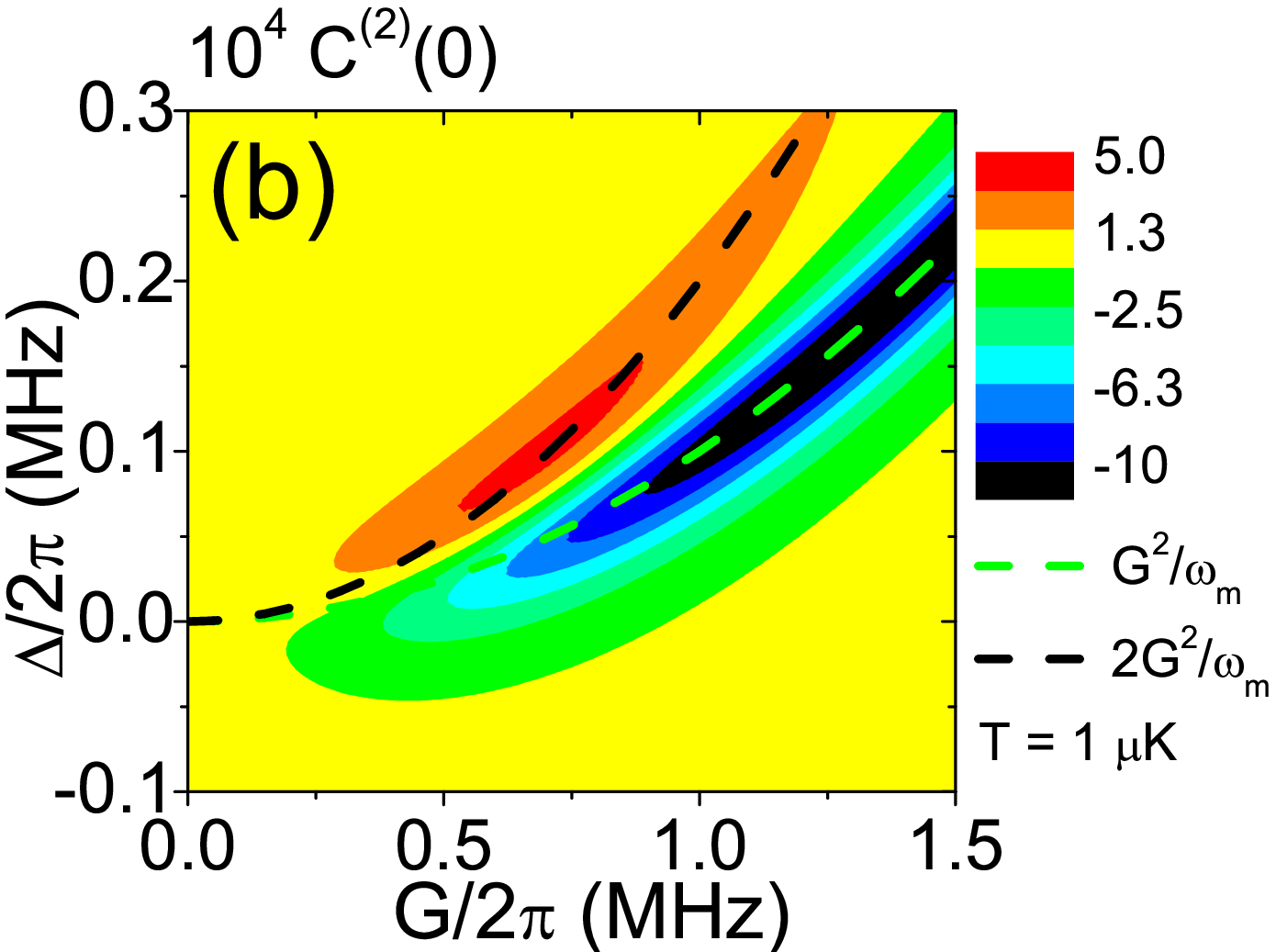}
\includegraphics[bb=0 0 375 275, width=4.1 cm, clip]{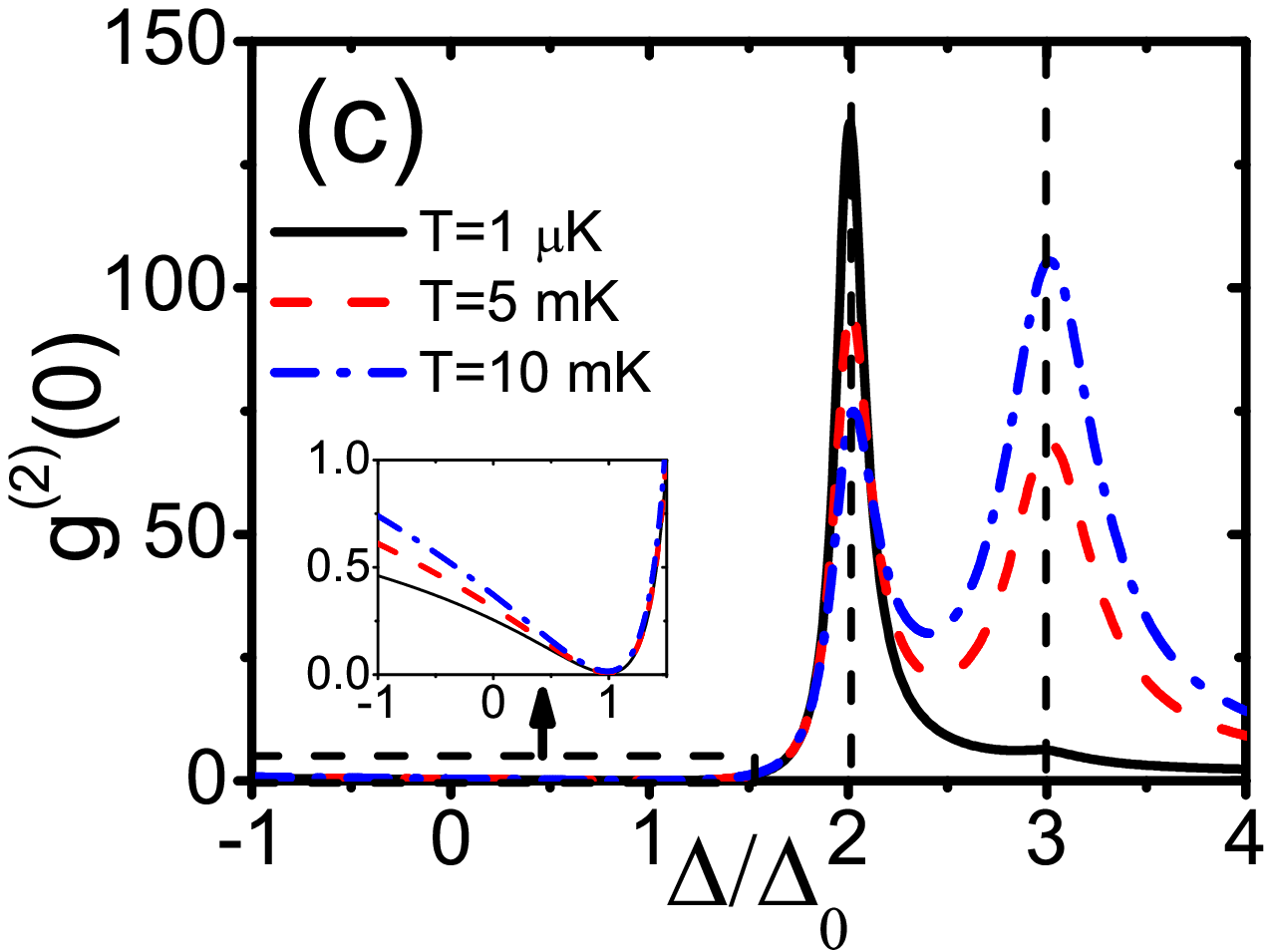}
\includegraphics[bb=0 0 375 275, width=4.1 cm, clip]{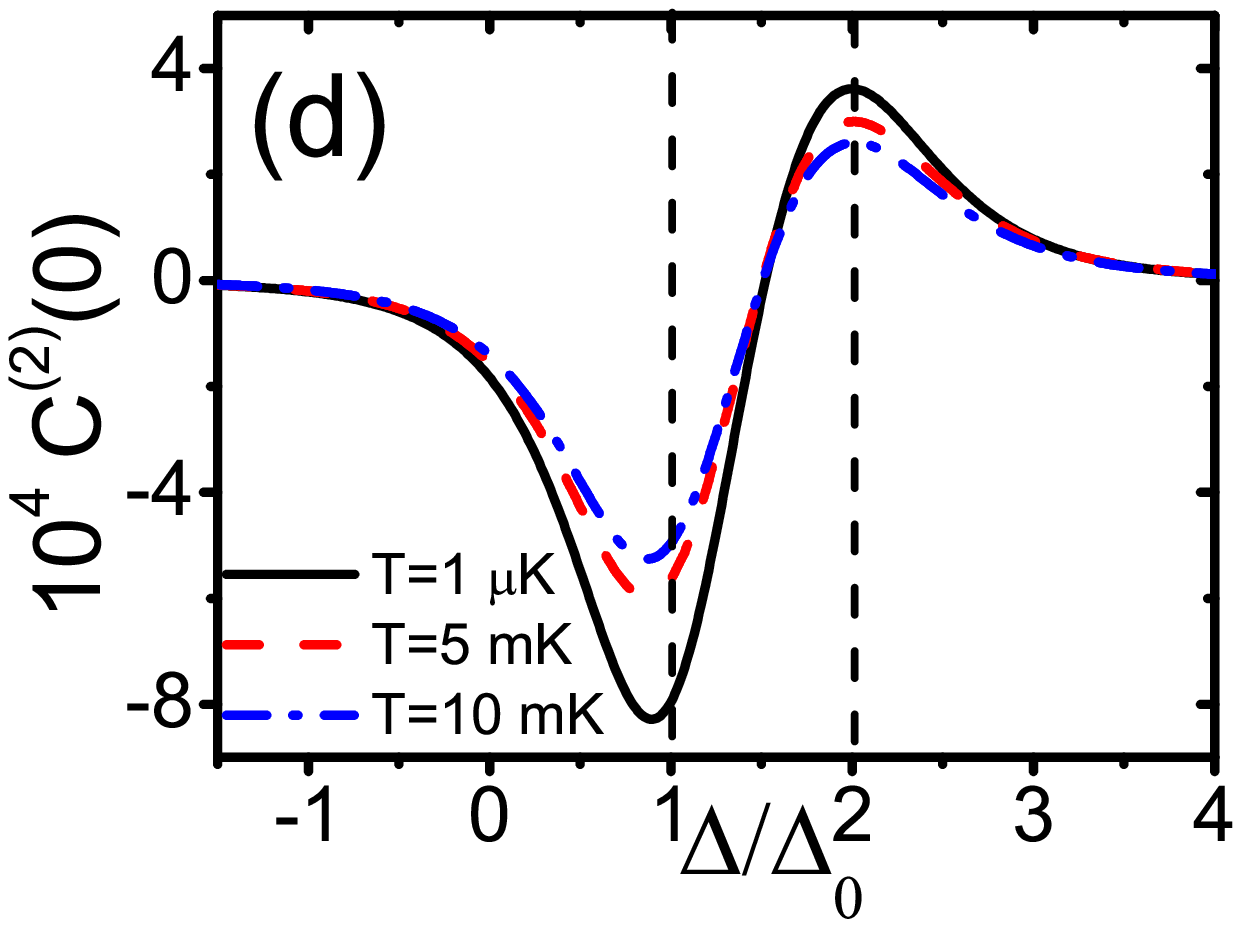}
\caption{(Color online) $g^{\left( 2\right) }\left( 0\right)$  and
$C^{\left( 2\right) }\left( 0\right)$ versus
$\Delta/\Delta_{0}$ and $G$ for $T=1$ $\mu$K in (a) and (b),
respectively. $g^{\left( 2\right) }\left( 0\right)$ versus
$\Delta/\Delta_{0}$ for different $T$ at $G/2\pi=2.5$ MHz in (c).
$C^{\left( 2\right) }\left( 0\right)$ versus
$\Delta/\Delta_{0}$ for different $T$ at $G/2\pi=0.75$ MHz in (d).
Other parameters are $\varepsilon_{c}/2\pi=0.01$ MHz,
$\gamma/2\pi=0.1$ MHz, $\omega_{m}/2\pi=10$ MHz, and
$\gamma_{m}/2\pi=0.01$ MHz.} \label{fig2}
\end{figure}

To explore two-photon tunneling and compare it with the photon
blockade, $g^{(2)}(0)$ and
$C^{(2)}(0)$ are plotted in Figs.~\ref{fig2}(a) and 2(b) as the functions of $\Delta$ and $G$ for given
$T$ in the steady state. Each figure has two curves, corresponding to single-
($\Delta=\Delta_{0}$) and two-photon ($\Delta=2\Delta_{0}$) resonant excitations from the
ground state to the first- and the second-excited states of the
cavity field, respectively. Figures~\ref{fig2}(a) and 2(b) clearly show
$g^{(2)}(0)<1$ and $C^{(2)}(0)<0$ for the single-photon resonant
excitation when $G$ is bigger than $\gamma$, that is, the
single-photon phenomenon or photon blockade occurs as shown in
Refs.~\cite{Rabl,Nunnenkamp}. However, when the frequency
$\omega_{c}$ of the probe field equals half of the transition
frequency from the ground state to the second excited state, i.e.,
$\Delta=2\Delta_{0}$, we find $g^{(2)}(0)>1$ and $C^{(2)}(0)>0$ as
shown in Figs.~\ref{fig2}(a) and 2(b). This means that the
single-photon transition from the ground state to the first excited
state is suppressed, but the second photon can enter the cavity and
make resonant transition from the ground state to the second
excited state together with the first photon. That is, the photon-induced
tunneling happens and photons can be absorbed in pairs
simultaneously.

\begin{figure}
\includegraphics[bb=5 230 590 705, width=8 cm, clip]{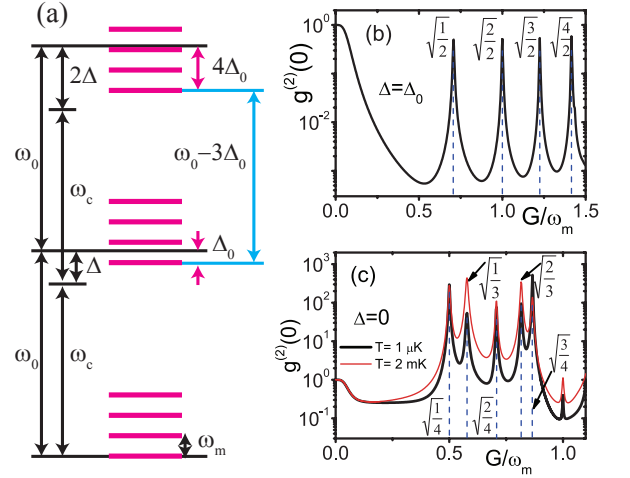}
\caption{(Color online) (a) Energy level diagram of the
optomechanical system. The horizontal long black lines linked
by the vertical black lines denote the energy levels of the cavity
field with frequency $\omega_{0}$. The
horizontal bright purple lines denote the energy levels corresponding to
the states $|n,\widetilde{m}\rangle$. The horizontal long cyan lines linked
by the vertical cyan line with two arrows denotes the transition of
$|1,\widetilde{m}\rangle \rightarrow |2,\widetilde{m}\rangle$. (b)
$g^{(2)}(0)$ versus $G/\omega_{m}$ for
$\Delta=\Delta_{0}$, $T=1$ $\mu$K. (c) $g^{(2)}(0)$ versus $G/\omega_{m}$ for $\Delta=0$, $T=1$ $\mu$K (black) and
$2$ mK (red). Other parameters are the same as in
Fig.~\ref{fig2}(c). } \label{fig3}
\end{figure}

We further show the effect of $T$ on the photon tunneling via
$g^{(2)}(0)$ and $C^{(2)}(0)$ in Figs.~\ref{fig2}(c) and 2(d). We
find that the low $T$ makes photon-induced tunneling and photon
blockade easily observed. It is easily understood because
the higher $T$ corresponds to the bigger decay rate $\gamma$ of the
cavity field as shown in Fig.~\ref{fig1}(b). Thus, with the increase
of $T$, $g^{(2)}(0)$ and $C^{(2)}(0)$ increase under the
single-photon ($\Delta=\Delta_{0}$) resonant driving, but decrease
under the two-photon ($\Delta=2\Delta_{0}$) resonant driving.
Moreover, there is an additional peak in Fig.~\ref{fig2}(c) for
$g^{(2)}(0)$ at $\Delta=3\Delta_{0}$ corresponding to the transition
between two eigenstates $|1,\widetilde{m}\rangle$ and
$|2,\widetilde{m}\rangle$ as shown in Fig.~\ref{fig3}(a). We find
that the height of the resonant peak at $\Delta=3\Delta_{0}$
increases when $T$ becomes higher. Qualitatively, this is because
$T$ changes the population distribution, especially it enhances the
population in higher energy levels, and then the transition from
$|1,\widetilde{m}\rangle$ to $ |2,\widetilde{m}\rangle$ is also
enhanced. All this results in the increase of $g^{(2)}(0)$ at
$\Delta=3\Delta_{0}$ with the increase of $T$.

The effect of the phonon states on the photon blockade has been
studied in Refs.~\cite{Rabl,liao2}. It was found that the two-photon
resonant transition between $|0,\widetilde{0}\rangle$ and
$|2,\widetilde{m}\rangle$ can also occur when $\Delta_{0}=m
\omega_{m}/2$ under the photon blockade condition
$\Delta=\Delta_{0}$. Thus there are resonant peaks located at
$G/\omega_{m}=\sqrt{m/2}$ for $g^{(2)}(0)$ with different phonon
states $|\widetilde{m}\rangle$ (e.g., $m=1,2,3,4$) as shown in
Fig.~\ref{fig3}(b). Similarly, we find that the phonon states also
affect the photon-induced tunneling. For example, Fig.~\ref{fig3}(c)
shows five peaks for $g^{(2)}(0) \gg 1$ in the regime $0 < G
<\omega_{m}$ under the condition $\Delta=0$ (i.e.,
$\omega_{0}=\omega_{c}$). As shown Fig.~\ref{fig3}(a), these five
peaks correspond to two types of the resonant conditions: (i) if
$\Delta_{0}=m \omega_{m}/4$, then the resonant transition between
$|0,\widetilde{0}\rangle$ and $|2,\widetilde{m}\rangle$ can occur
with the peaks at $G/\omega_{m}=\sqrt{m/4}$ ($m=1,2,3$);
(ii) when $\Delta_{0}=m \omega_{m}/3$, another resonant transition
between $|1,\widetilde{0}\rangle$ and $|2,\widetilde{m}\rangle$ is
allowed with the peaks at $G/\omega_{m}=\sqrt{m/3}$ ($m=1,2$). As the transitions
$|1,\widetilde{0}\rangle\rightarrow |2,\widetilde{m}\rangle$ are enhanced with the increase of
$T$, the peak at $G/\omega_{m}=\sqrt{m/3}$ ($m=1,2$) increases
when $T$ is increased.

To approximately obtain the conditions of the photon blockade and
the photon-induced tunneling, we analyze the relation between
$g^{(2)}(0)$ and the probabilities $P(n)$ of $n$-photon
distribution. $P\left( n\right)$ corresponding to the state
in Eq.~(\ref{eq:7}) can be given by $P\left( n\right) ={\sum
}_{m}\rho_{n,m;n,m}(t)$; then we have
\begin{equation}\label{eq:11}
g^{\left( 2\right) }\left( 0\right) =\frac{{\rm Tr} [\rho a^{\dag
2}a^{2}] }{ [{\rm Tr} (\rho a^{\dag }a)]^2}=%
\frac{\underset{n}{\sum }n\left( n-1\right) P\left( n\right) }{\left[
\underset{n}{\sum }nP\left( n\right) \right]^{2}}.
\end{equation}
In the limit of the weak probe field (e.g.,
$\varepsilon_{c}=0.1\gamma$), $P\left( n\right) \gg P\left(
n+1\right)$, $g^{(2)}(0)$ can be given approximately by
\begin{eqnarray}\label{eq:12}
g^{\left( 2\right) }\left( 0\right) &\approx & \frac{2P\left(
2\right) }{[P\left( 1\right)]^2}
\approx \left\vert \frac{\gamma +i 2\left( \Delta -\Delta _{0}\right) }{%
\gamma +i2\left( \Delta -2\Delta _{0}\right) }\right\vert ^{2},
\end{eqnarray}
in the steady state. Here we have assumed that the phonon is in its
ground state to obtain the second approximated expression. This
assumption has also been made in the following derivations in
Eq.~(\ref{eq:13}) and Eq.~(\ref{eq:15}). Equation~(\ref{eq:12}) means
that $g^{(2)}(0)$ is proportional to the ratio between the
probability to prepare a two-photon state and that to prepare two
single-photon states independently. In Fig.~\ref{fig4}(a),
$g^{(2)}(0)$ and its approximated expression in Eq.~(\ref{eq:12})
versus the normalized detuning $\Delta$ are numerically simulated in
the steady state. We find that the approximated expression fits well
with the exact solution of $g^{(2)}(0)$ in the limit of the weak
probe field. Thus $g^{(2)}(0)>1$ [or $g^{(2)}(0)<1$] means that the
probability to excite the two-photon state is bigger (smaller) than that
to excite two single-photon states independently, and then
photon-induced tunneling (photon blockade) happens. To
quantitatively show the relation between the height of
Fig.~\ref{fig2}(c) [and also Fig.~\ref{fig4}(a)] and system
parameters, we derive an approximated expression,
\begin{equation}
g^{\left( 2\right) }\left( 0\right) \approx (\gamma
^{2}+4\Delta _{0}^{2})/\gamma ^{2},
\end{equation}
from Eq.~(\ref{eq:12}) with $\Delta =2\Delta _{0}$, which
corresponds to the height of the peak in Fig.~\ref{fig4}(a) and
Fig.~\ref{fig2}(c) at $\Delta=2\Delta_{0}$. Thus the height of the
peak in Fig.~\ref{fig4}(a) and Fig.~\ref{fig2}(c) is approximately
determined by the ratio between  $\gamma$ and $\Delta_{0}$. We can
also obtain
\begin{equation} g^{\left(
2\right) }\left( 0\right) \approx \gamma ^{2}/(\gamma
^{2}+4\Delta _{0}^{2}),
\end{equation}
for $\Delta =\Delta _{0}$, which corresponds to the deep in
Fig.~\ref{fig4} (a) and Fig.~\ref{fig2}(c). It is clear that the
condition to observe photon blockade or photon-induced tunneling is
$ \Delta _{0} \gg \gamma/2$.

\section{Multi-photon tunneling}

\begin{figure}
\includegraphics[bb=35 5 370 270, width=4.1 cm, clip]{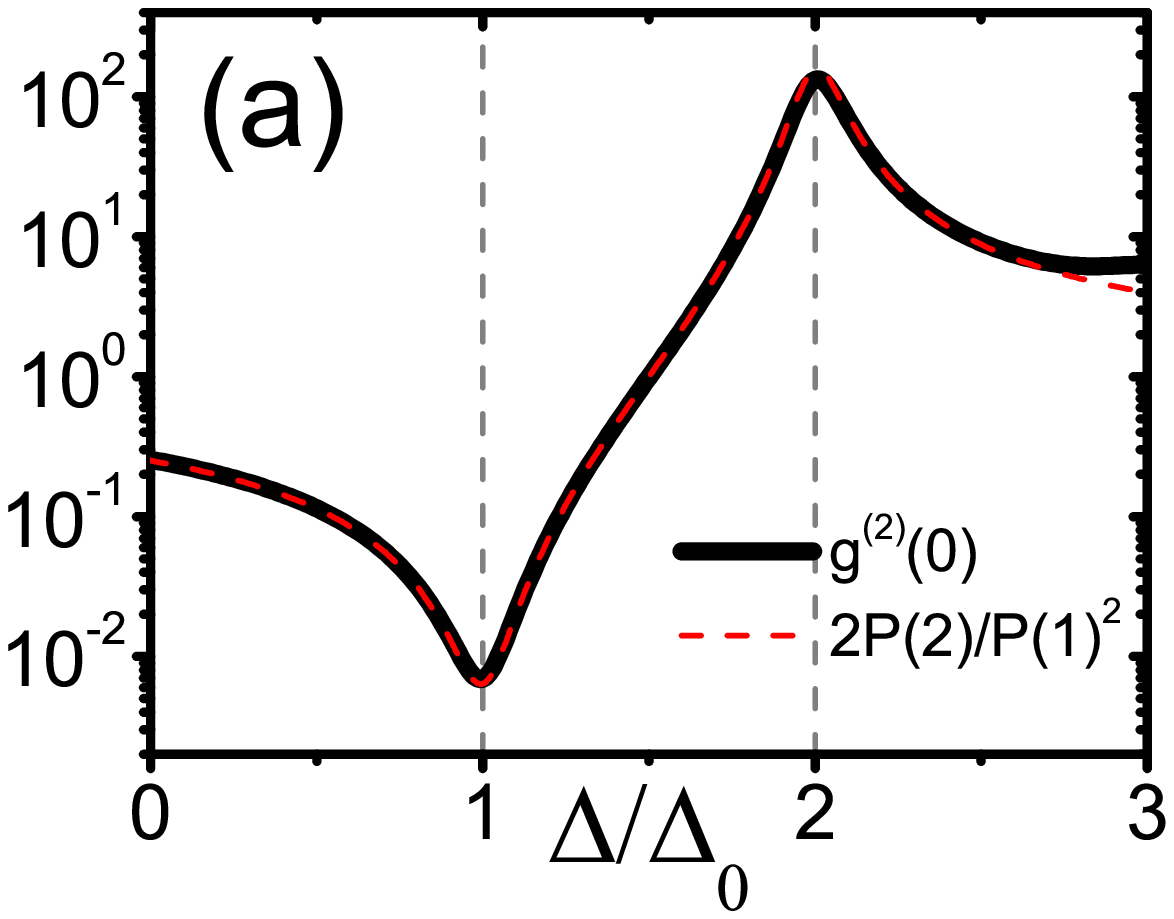}
\includegraphics[bb=35 5 370 270, width=4.1 cm, clip]{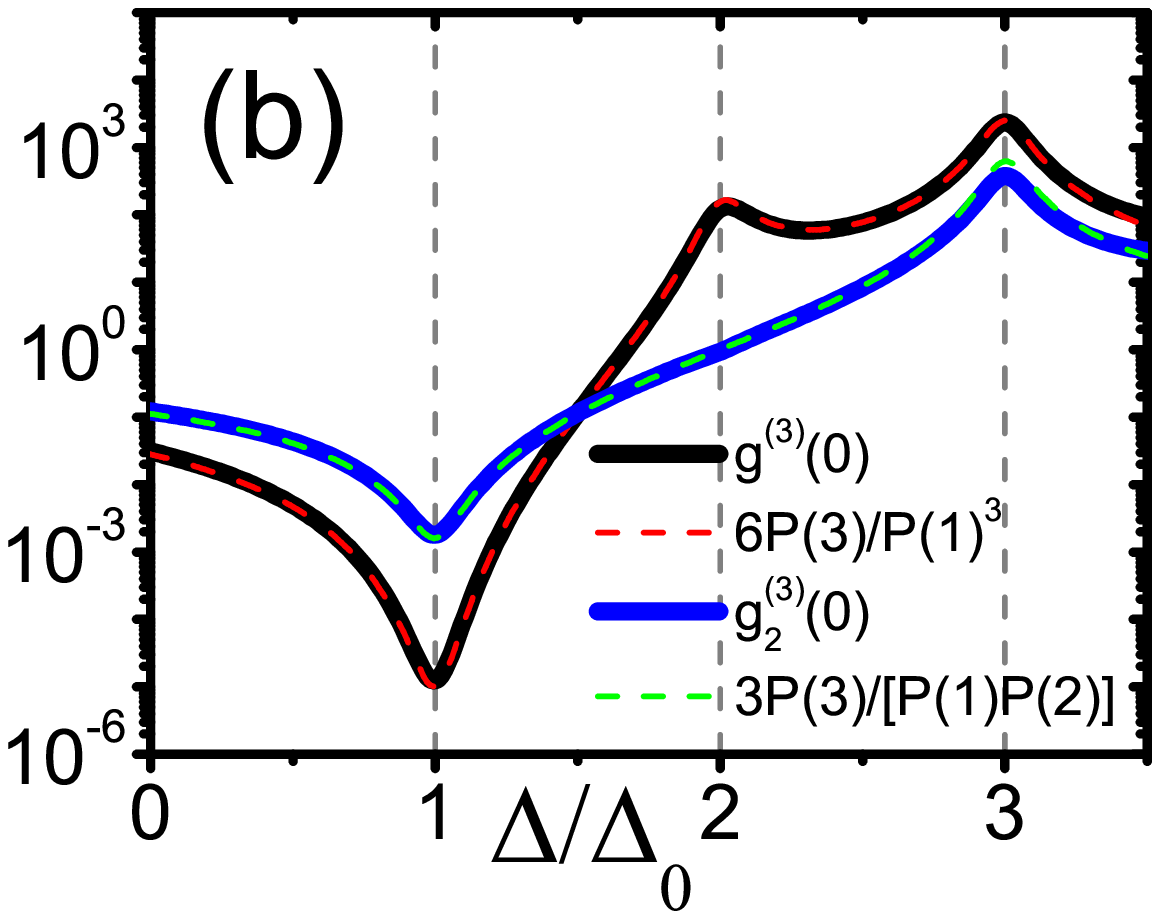}
\caption{(Color online) (a) $g^{(2)}(0)$ and $2 P(2)/P(1)^2$, (b)
$g^{(3)}(0)$, $6P(3)/P(1)^3$, $g^{(3)}_{2}(0)$ and
$3P(3)/[P(1)P(2)]$  plotted as functions of the normalized detuning
$\Delta/\Delta_{0}$ for $G=2.5$ MHz and $T=1$ $\mu$K. The other
parameters are the same as in Fig.~\ref{fig2}(c).} \label{fig4}
\end{figure}

In principle, the multiphoton tunneling (more than two) can be
studied via Eq.~(\ref{eq:5}). However, for the sake of simplicity,
we just use the three-photon case as an example to show multiphoton
tunneling. Using the similar discussions as for Eq.~(\ref{eq:11})
and Eq.~(\ref{eq:12}), $g^{\left( 3\right) }\left( 0\right)$ can be
approximately written as
\begin{eqnarray}\label{eq:13}
g^{\left( 3\right) }\left( 0\right) &\equiv & \frac{{\rm Tr}\left(\rho
a^{\dag 3}a^{3}\right)}{[{\rm Tr}\left(\rho a^{\dag }a\right)]
^{3}} \approx \frac{6P\left( 3\right) }{\left[ P\left( 1\right)
\right] ^{3}} \\
&\approx &\left\vert \frac{\left[\gamma +i 2\left( \Delta -\Delta_{0} \right) \right] ^{2}}{\left[ \gamma +i2\left(
\Delta -2\Delta_{0}\right) \right] \left[ \gamma +i2\left(
\Delta -3\Delta_{0}\right) \right] }\right\vert ^{2} \nonumber
\end{eqnarray}
in the steady state with the weak probe field. It is clear that
$g^{(3)}(0)$ is proportional to the ratio between the probability to
create a three-photon state and that to create three single photons
independently. If we introduce a quantity
\begin{equation}
g_{2}^{\left( 3\right) }\left( 0\right) \equiv\frac{g^{\left(
3\right) }\left( 0\right)}{g^{\left( 2\right) }\left(
0\right)}=\frac{{\rm Tr}\left(\rho a^{\dag 3 }a^3\right) }{{\rm
Tr}\left(\rho a^{\dag }a \right) {\rm Tr}\left(\rho a^{\dag
2}a^{2}\right)}
\end{equation}
to characterize the ratio between the normalized third-order and
second-order correlation functions, then we can obtain
\begin{eqnarray}\label{eq:15}
g_{2}^{\left( 3\right) }\left( 0\right) &\approx& \frac{3P\left(
3\right) }{P\left( 1\right) P\left( 2\right) }\approx \left\vert
\frac{\gamma +i 2\left( \Delta -\Delta_{0}\right) }{
\gamma+i2\left( \Delta -3\Delta_{0}\right) }\right\vert ^{2},
\end{eqnarray}
in the steady state. $g_{2}^{\left( 3\right) }\left( 0\right)$ is
proportional to the ratio between the probability to create a
three-photon state and the joint probability to create a single-
plus a two-photon state independently, which is another way to
create three photons.

In Fig.~\ref{fig4}(b),  $g^{(3)}(0)$ and its approximated expression
$6P(3)/[P(1)]^3$ as well as  $g^{(3)}_{2}(0)$ and its approximated
expression $3P(3)/[P(1)P(2)]$ are plotted as functions of
$\Delta/\Delta_{0}$. We find that the approximation in both
Eq.~(\ref{eq:13}) and Eq.~(\ref{eq:15}) is valid for the weak probe
field. Under the condition that $\Delta _{0}\gg \gamma/2$, we can
further find
\begin{equation}\label{eq:17}
g^{\left( 3\right) }\left( 0\right)  \approx \left(\gamma ^{2}/
8\Delta _{0}^{2}\right) ^{2}, \quad
g_{2}^{\left( 3\right) }\left( 0\right)  \approx (\gamma/
4\Delta _{0})^{2},
\end{equation}
for the resonant condition $\Delta =\Delta _{0}$, and
\begin{equation}\label{eq:18}
g^{\left( 3\right) }\left( 0\right)  \approx (2\Delta _{0}/
\gamma)^{2},  \quad
g_{2}^{\left( 3\right) }\left( 0\right)  \approx 1,
\end{equation}%
for the resonant condition $\Delta =2\Delta _{0}$, as well as
\begin{equation}\label{eq:19}
g^{\left( 3\right) }\left( 0\right)  \approx (8\Delta _{0}/\gamma)^{2},  \quad
g_{2}^{\left( 3\right) }\left( 0\right)  \approx (4\Delta _{0}/
\gamma)^{2},
\end{equation}
for the resonant condition $\Delta =3\Delta _{0}$.
Equations~(\ref{eq:18}) and (\ref{eq:19}) show that the heights of the resonant
peaks are approximately determined by the ratio of $\Delta_{0}$ and
$\gamma$ as for two-photon tunneling.

Figure~\ref{fig4}(b) and Eqs.~(\ref{eq:17})-(\ref{eq:19}) show the following. (i) For the single-photon
resonant excitation ($\Delta=\Delta_{0}$), $g^{(3)}(0)<1$ and
$g_{2}^{\left( 3\right) }\left( 0\right)<1$, which means that
the photon inside the cavity is antibunching or the subsequent
photons will be blocked by the photons inside the cavity. (ii) For
the three-photon resonant excitation ($\Delta=3\Delta_{0}$),
$g^{(3)}(0)>1$ and $g_{2}^{\left( 3\right) }\left( 0\right)>1$,
that is, the photon inside the cavity is bunching or the cavity
can absorb three photons simultaneously. (iii) For the
two-photon resonant excitation ($\Delta=2\Delta_{0}$),
$g^{(3)}(0)>1$ which means that the probability of three-photon
absorption is bigger than that of three single-photon absorption
independently. We also find $g_{2}^{\left( 3\right) }\left(
0\right)\approx 1$ at the point $\Delta=2\Delta_{0}$, which means
that the joint probability of the photon absorption in pairs after or before single-photon
absorption is approximately equal to that of three-photon
absorption by the cavity. Therefore, the necessary condition for
absorbing three photons by the cavity simultaneously (or
three-photon tunneling) is $\Delta=3\Delta_{0}$. In this
condition, $g^{(3)}(0)>1$ and $g_{2}^{\left( 3\right) }\left(
0\right)>1$, the probability of generating three photons
simultaneously is bigger than those of generating three
single-photons independently and generating a single-photon after
or before generating a photon pair.

\section{Conclusions}

In summary, we have studied the photon induced tunneling and compared
it with the photon blockade in optomechanical systems with the
strong optomechanical coupling. Our study shows that the cavity
field can exhibit photon antibunching when the probe field is
resonant with the transition from the ground state to the first
excited state of the cavity field. However, the two-photon tunneling
occurs when the frequency of the probe field equals half of
the transition frequency from the ground state to the second excited
state of the cavity field. Moreover, we find that three-photon
tunneling occurs when the frequency of the probe field satisfies the
condition of the three-photon resonant excitation from the ground
state to the third excited state. We further show that the phonon
states greatly affect the multiphoton resonance in the certain
condition. Our studies can be easily generalized to the $n$-photon
($n>3$) case. Our results show that the photon-induced tunneling can
be experimentally observed when the optomechanical system approaches
the strong-coupling limit.

\emph{Note added.} We note a related paper that appeared recently~\cite{Kronwald}.

\emph{Acknowledgement.} Y.X.L. is supported by the National Natural Science Foundation of
China under Grants No. 10975080 and No. 61025022.

\bibliographystyle{apsrev}
\bibliography{ref}

\end{document}